# Enhancement of magnetoelectric coupling in Cr doped $Mn_2O_3$

**Mohit Chandra[1], Satish Yadav[1], Rajeev Rawat[1], Kiran Singh[1,2*]**

[1]UGC-DAE Consortium for Scientific Research, University Campus, Khandwa Road Indore-452001, India
[2]Department of Physics, Dr. B. R. Amdedkar National Institute of Technology, Jalandhar, 144011, India

**Abstract**

The effect of Cr doping with nominal compositions $Mn_{2-x}Cr_xO_3$ ($0 \leq x \leq 0.10$) has been undertaken to investigate its effect on structural, magnetic, dielectric and magnetoelectric properties. The Cr doping transformed the room temperature crystal structure from orthorhombic to cubic symmetry. Similar to α-$Mn_2O_3$, two magnetic transitions have been observed in the Cr doped samples. The effect of Cr doping is significant on the low temperature transition i.e. the lower magnetic transition shifted towards higher temperature (25 K for pristine to 40 K for x=0.10) whereas the high temperature transition decreases slightly with increasing Cr content. A clear frequency independent transition is observed in complex dielectric measurements for all compositions around high temperature magnetic ordering. Interestingly, the magnetodielectric behaviour enhanced enormously ~21% with Cr substitution as compared to pristine $Mn_2O_3$.

[*]**Corresponding Author:**

Dr. Kiran Singh

Department of Physics, Dr. B. R. Ambedkar National Institute of Technology, Jalandhar-144011, India

**Email:** singhkp@nitj.ac.in

kpatyal@gmail.com

Magnetoelectric (ME) materials based on the direct combination of electric and magnetic order parameters in a single phase system attracted a great attention because of their interesting physical properties as well as their potential in different applications such as memory devices, sensors, spintronic etc. [1-5]. The presence of ME coupling is technologically important for the control of magnetic properties with an electric field and vice-versa. However, the coexistence of ferroelectricity and magnetism in a single phase material are rare because of the different microscopic mechanism for ferroelectricity and ferromagnetism [5-7]. In last decades, many multiferroic compounds such as $MnWO_4$, $BaMnO_3$, $GeV_4S_8$, $Ba_2FeSbSe_5$, $AgCrS_2$ and $CdCr_2S_4$, etc. [8-13] has been found and lots of studies have been focused on the search of new multiferroic materials and understanding their microscopic origin.

The $ABO_3$ type compound are one of the most studied class of oxide materials for their multiferroicity and ME coupling [14-16]. Stoichiometrically, the $Mn_2O_3$ is same as the first ME material $Cr_2O_3$. The $Mn_2O_3$ has different phases depending upon the synthesis conditions. Recently, we have explored the multiferroics and ME coupling in α-$Mn_2O_3$ [14]. Cong et al. also observed spin induced multiferroicity in the high temperature and high pressure phase ξ-$Mn_2O_3$ [17]. The pristine α-$Mn_2O_3$ is orthorhombic at room temperature [18, 14] and undergoes cubic to orthorhombic transition at 308 K [19,20]. The effect of substitution has interesting effects on magnetoelectric properties [21-24]. In this system, Mn formed octahedral surrounding and having Jahn-Teller (J-T) active cation. The substitution of Cr at Mn site which is non-active J-T cation can influence the structural, magnetic, dielectric and magnetodielectric properties of α-$Mn_2O_3$ because of the empty $e_g$ orbital in case of $Cr^{3+}$ than one electron in $e_g$ level in case of $Mn^{3+}$. Here, we present the effect of Cr doping in α-$Mn_2O_3$ on its structural, magnetic, dielectric and magnetodielectric properties. The Cr doping with x=0.10 shows highest magnetodielectric effect which is around 21% higher than the pristine α-$Mn_2O_3$.

The polycrystalline sample of $Mn_{2-x}Cr_xO_3$ (0 ≤ x ≤ 0.10) has been prepared through standard solid state reaction method. High purity $MnO_2$ and $Cr_2O_3$ (99.99%) have been mixed in stoichiometric ratio and sintered at 800°C for 12 h. The room temperature (RT) synchrotron x-ray diffraction (SXRD) measurements were performed at beamline-12 at Indus-2, RRCAT Indore (λ=0.798600 Å) [25]. The dc magnetization measurements were done on a 7 T Quantum Design SQUID-VSM. The temperature and magnetic field dependent complex dielectric measurements were performed using a home-made insert coupled with an 9 T superconducting magnet from American Magnetics and a Keysight E4980A LCR meter [14]. The temperature dependent remnant electric polarization is performed using Keithley 6517B electrometer in Columbic mode in different magnetic fields during warming at a rate of 2 K/min. The direction of electric and magnetic fields was parallel to each other. In case of dielectric and electric polarization under magnetic field, the magnetic field was applied at the lowest temperature.

The RT Rietveld refinement of SXRD pattern for $Mn_{2-x}Cr_xO_3$ ($0.01 \leq x \leq 0.10$) is shown in Fig. 1. The refinement is performed using Fullprof software [26]. Our structural analysis infers that for Cr doped samples the crystal structure is cubic at RT with space group *Ia-3* instead of orthorhombic as in case of pristine α-$Mn_2O_3$. The insets in Fig. 1 (a-c) show the zoomed view of diffraction pattern for 2θ=23° to 26°. It is interesting to note that in Cr doped samples, there is no additional peak at 2θ~24.30° corresponding to *hkl* ~ (034) as observed for α-$Mn_2O_3$ in our earlier work [14] within the instrument resolution. These results suggest that the crystal structure of Cr doped samples is cubic at RT. Earlier Geller et al. also shows that Cr doping leads to the formation of cubic structure at RT [20]. The α-$Mn_2O_3$ undergoes cubic to orthorhombic transition at 308 K. These results corroborate that Cr doping lower the cubic to orthorhombic transition. The change of crystal structure from orthorhombic to cubic can be explained on the basis of replacement of J-T active cation ($Mn^{3+}$) with non J-T active cation ($Cr^{3+}$) which reduces the structural distortion. Even 1% Cr doping shows cubic structure at RT. The arrow in the insets of Fig. 1(a-c) show the 2θ value where the peak related to orthorhombic structure has been observed in pure $Mn_2O_3$. The details of unit cell for $Mn_{2-x}Cr_xO_3$ ($0.01 \leq x \leq 0.10$) is given in Table-I. Table-I suggests that the lattice parameters as well as the unit cell volume decrease with increasing Cr-content. The lattice parameters changes systematically with increasing x, which imply that Cr-ions do replace the Mn-ions. The schematic representation of the crystalline structure of $Mn_{1.9}Cr_{0.10}O_3$ on the basis of SXRD results is presented in Fig.1 (d).

Temperature dependent dc magnetic susceptibility (χ=M/H) at 500 Oe during zero field cooled warming (zfc), field cooled cooling (fcc) and field cooled warming (fcw) conditions for $Mn_{2-x}Cr_xO_3$ ($0.01 \leq x \leq 0.10$) is presented in Fig. 2 (a), (d) and (g). These plots show that for x=0.01, 0.05 and 0.10; χ exhibit two antiferromagnetic transitions ($T_{N1}$) ~79, 78, 77 K and other transition ($T_{N2}$) around 37, 35 and 40 K, respectively. For x=0.10, an additional clear feature is observed around 94 K. These transitions are clearly seen in the dχ/dT *vs* T plots and presented in the Fig. 2 (c), (f) and (i) (left y-axis). As compared to pure α-$Mn_2O_3$, $T_{N1}$ decreases slightly whereas the $T_{N2}$ shifts towards higher temperature which is consistent with earlier report [20]. In addition all χ vs T plots show that the χ increase rapidly after $T_{N2}$ which is different from pristine $Mn_2O_3$. These observations suggest the modifications in the magnetic interactions with Cr doping.

The fcc and fcw curves for all the compositions show a thermal hysteresis around lower magnetic transition as seen for pure $Mn_2O_3$, which suggest the *first order* nature of this transition. The effective magnetic moment ($\mu_{eff}$) and the Curie-Weiss paramagnetic temperature ($\theta_P$) are determined by Curie-Weiss law $\chi = \frac{C}{(T-\theta_P)}$ (where $C$ is Curie-Weiss constant) from the linear part of the inverse dc susceptibility data at high temperature (160-300 K), not shown here (see supplementary information Fig. S1). The observed results are summarized in Table-II. From table-II it is clear that the value of $\mu_{eff}$ decreases slightly with increasing Cr content. For α-$Mn_2O_3$ sample, $\mu_{eff}$ =4.87$\mu_B$ which is approximately equal to the theoretical value of 4.90$\mu_B$ for $Mn^{3+}$. The valence state of Cr should be +3, the spin only value is calculated using equation

$$\mu_{eff.} = \sqrt{(1-x)\mu_{Mn}^2 + x\mu_{Cr}^2}$$ where x is the Cr ion content. The experimentally extracted and theoretical value of $\mu_{eff}$ for different Cr content is given in Table II. The magnitude of $\theta_P$ is very high as compared to the pristine sample, which further suggests the modifications in magnetic interactions with Cr doping. There is no significant change in the MH behavior of all these samples and show linear behavior without any hysteresis which again confirm the antiferromagnetic ground state of the studied samples (see supplementary information Fig. S2). We have not observed any magnetic field induced transition at least up to 70 kOe magnetic field. This modification in magnetic properties is due to the replacement of J-T active cation $Mn^{3+}$ with non J-T active cation $Cr^{3+}$. In case of $Mn^{3+}$ there is one electron in $e_g$ level whereas it is empty in case of $Cr^{3+}$. The Cr doping leads to complex magnetic interactions with different combinations [27]. The possible magnetic interactions are Mn-O-Mn, Mn-O-Cr and Cr-O-Cr.

The temperature dependent dielectric constant of $Mn_{2-x}Cr_xO_3$ (0.01 ≤ x ≤ 0.10) at 100 kHz at 0 and 80 kOe magnetic fields during warming (1K/min) are shown in Fig. 2 (b), (e) and (h). The dielectric constant decreases with decreasing temperature and exhibit a frequency independent transition around 79, 78 and 77 K for 0.01, 0.05 and 0.10, respectively. The transition in dielectric results is frequency independent as in case of pristine $Mn_2O_3$ [14]. The dielectric behavior at different frequencies for x=0.01 and 0.10 is presented in the inset of the Fig. 2 (b) and (h), respectively. The decrease in the transition temperature in dielectric results with Cr content is in accordance with the magnetic transition. The occurrence of dielectric anomaly at magnetic ordering illustrates the coupling between magnetic and dielectric properties. Moreover, the dielectric anomalies at $T_{N1}$ for all compositions are magnetic field independent as in case of pure $Mn_2O_3$. In addition, below 20 K the dielectric constant increases with decreasing temperature in both the 0 and 80 kOe fields. The change in the dielectric value *vs* temperature is very small and we could not able to detect any clear change in dielectric around $T_{N2}$. To see the changes across the magnetic transitions, we have plotted $d\varepsilon'/dT$ *vs* T for all compositions and presented in Fig. 2 (c, f and i; right y-axis). These figures show that there is small change in the slope of $d\varepsilon'/dT$ around 20 K. The temperature dependent $d\chi/dT$ and $d\varepsilon'/dT$ has one to one correspondence. Fig. 2 (c,f and i) corroborate that like pristine $Mn_2O_3$ there is a tail like feature above $T_{N1}$ for all the studied compositions and this feature is more prominent in case of x=0.10. The value of tan$\delta$ at for x= 0.01, 0.05 and 0.10 are also very small shown, see supplementary information Fig. S3.

To investigate the effect of magnetic field on the dielectric behavior, we have performed isothermal magnetodielectric (MD) measurements at different temperatures and frequencies for different Cr content. The MD is calculated using relation $MD = (\varepsilon'_H - \varepsilon'_{H=0})/\varepsilon'_{H=0}$. The MD% for different Cr content at 5 K is shown in Fig. 3 (a). The observed MD is negative and it decreases with increasing magnetic field up to 80 kOe at lower temperature. Above magnetic ordering the value of MD is nearly zero. For α-$Mn_2O_3$, maximum MD value at 5 K is 0.015% and positive in sign. In case of Cr doping, the sign of MD is negative and its magnitude increases with Cr-doping till x=0.10. For x=0.10, the maximum MD at 5 K is ~0.21% which is around 21% higher than the pristine sample. These variation of dielectric with magnetic field endorse the enhancement of ME coupling in Cr doped $Mn_2O_3$ samples. The change in MD sign also suggests the change in magnetic interactions with Cr doping at Mn sites [28]. The MD for x=0.10 is shown at two different temperatures in Fig. 3 (b). The MD at 5 K infer that the dielectric behavior decreases almost quadratically till 60 kOe and then it decreases very slowly with further increase in magnetic field. This quadratic variation of MD suggests the contributions of higher order magnetoelectric coupling [29] at lower magnetic field (~60 kOe for x=0.10). The temperature dependent MD effect for x=0.10 is extracted from temperature dependent dielectric at 0 and 80 kOe fields (from Fig. 2h). This also suggests the maximum MD at lowest temperature. We would also like to mention here that the Cr doping enhanced the dc resistivity and confirm that the studied samples are highly insulating. Therefore, the effect of any extrinsic contributions like magnetoresistance can be excluded.

To evident the ferroelectricity in Cr doped samples, temperature dependent remnant electric polarization (*P*) is measured under different magnetic fields. The electric polarization for x=0.10 under different magnetic fields at electric field ~217kV/m are presented in Fig. 4. The inset of Fig. 4 shows the polarization for x=0.05. This figure shows the emergence of *P* around 105 K for both compositions. The variation of *P* from low temperature to ~ 80 K is very small. However, there is sharp decrease in *P* across 80 K, which is near the magnetic ordering region. Moreover, the onset of *P* is coinciding with the first derivative of magnetization and dielectric as presented in the insets of Fig. 2 (c) and (i). This behavior is consisting with pristine $Mn_2O_3$ [14]. The *P* at different magnetic fields is also presented for x=0.10. The magnitude of *P* is suppressed with the application of magnetic field. Similar to the small anomalies in dielectric behavior around 20 K, cusp like feature is also observed in *P* around the same temperature. There is also small change in polarization around lower magnetic ordering. The one to one correspondence of *P* and magnetization results infers the coupling between electric and magnetic ordering.

In summary, we have studied the structural, magnetic and magnetoelectric properties of polycrystalline $Mn_{2-x}Cr_xO_3$ (0 ≤ x ≤ 0.10). Our structural analysis infers that the Cr doped samples crystallize in cubic symmetry with space group *Ia-3* at RT. The change in crystal structure from orthorhombic to cubic with Cr doping may be due to the reduction in the octahedral distortion due to the non J-T active Cr at Mn site. The effect of Cr doping is also observed on the magnetic and dielectric properties. The low temperature magnetic ordering shifted towards the higher temperature whereas the high temperature transition decreases slightly with increasing Cr content. The existence of dielectric transition and emergence of electric polarization near magnetic ordering temperature corroborates the correlation between the magnetic and dielectric properties in the studied samples. The significant enhancement in MD (~.21%) is observed for x=0.10 which is ~21% higher than the pristine $Mn_2O_3$. Our results demonstrate the enhancement of magnetoelectric properties in Cr doped $Mn_2O_3$.

**Acknowledgement:** The authors would like to thank Dr. R. J. Choudhary for the magnetic measurements. Authors are also thankful to Indus-2, beam line-12 members Dr. A.K. Sinha, Dr. Archana Sagdeo and Mr. M.N. Singh for synchrotron x-ray diffraction measurements.

**Table Captions:**

**Table I.** Lattice parameters and unit cell volume of $Mn_{2-x}Cr_xO_3$ (0.01 ≤ x ≤ 0.10) at room temperature.

**Table II.** Experimental and theoretical values of effective paramagnetic moment ($\mu_{eff}$) including $\theta_P$ for $Mn_{2-x}Cr_xO_3$ (0 ≤ x ≤ 0.10).

**Figure Captions:**

**Fig.1.** Rietveld refinement of SXRD patterns of (a) $Mn_{1.99}Cr_{0.01}O_3$ (b) $Mn_{1.95}Cr_{0.05}O_3$ and (c) $Mn_{1.9}Cr_{0.10}O_3$ at room temperature. (d) Schematic representation of the crystal structure for $Mn_{1.9}Cr_{0.10}O_3$.

**Fig.2** Temperature variation of magnetic susceptibility (a), (d) and (g); dielectric permittivity (b), (e) and (h) at 100 kHz at 0 and 80 kOe magnetic fields and inset in (e) and (h) shows the dielectric permittivity at different frequencies at 0 Oe. (c), (f) and (i) shows temperature variation of $d\chi/dT$ and $d\varepsilon'/dT$ for different compositions.

**Fig.3** (a) Isothermal magnetodielectric of different compositions of Cr doped $Mn_2O_3$ at 5 K. (b) Magnetodielectric for x=0.10 at 5 and 100 K. Inset shows the temperature dependent magnetodielectric for x=0.10 at 50 kHz extracted from temperature dependent dielectric at 0 and 80 kOe.

**Fig.4** Temperature dependent remnant polarization for x=0.10 at different magnetic fields, inset shows remnant polarization for x= 0.05.

Table I

| x | Lattice parameters(Å) | Volume(Å$^3$) |
|---|---|---|
| 0.01 | 9.4173(3) | 835.17(4) |
| 0.05 | 9.4036(4) | 831.54(7) |
| 0.10 | 9.3968(3) | 829.73(5) |

Table II

| x | $\theta_P(K)$ | $\mu_{eff}$ (Exp.) | $\mu_{eff}$ (Theo.) |
|---|---|---|---|
| 0 | -25.59 | 4.87 | 4.90 |
| 0.01 | -117.37 | 4.81 | 4.88 |
| 0.05 | -142.38 | 4.78 | 4.87 |
| 0.10 | -105.73 | 4.58 | 4.85 |

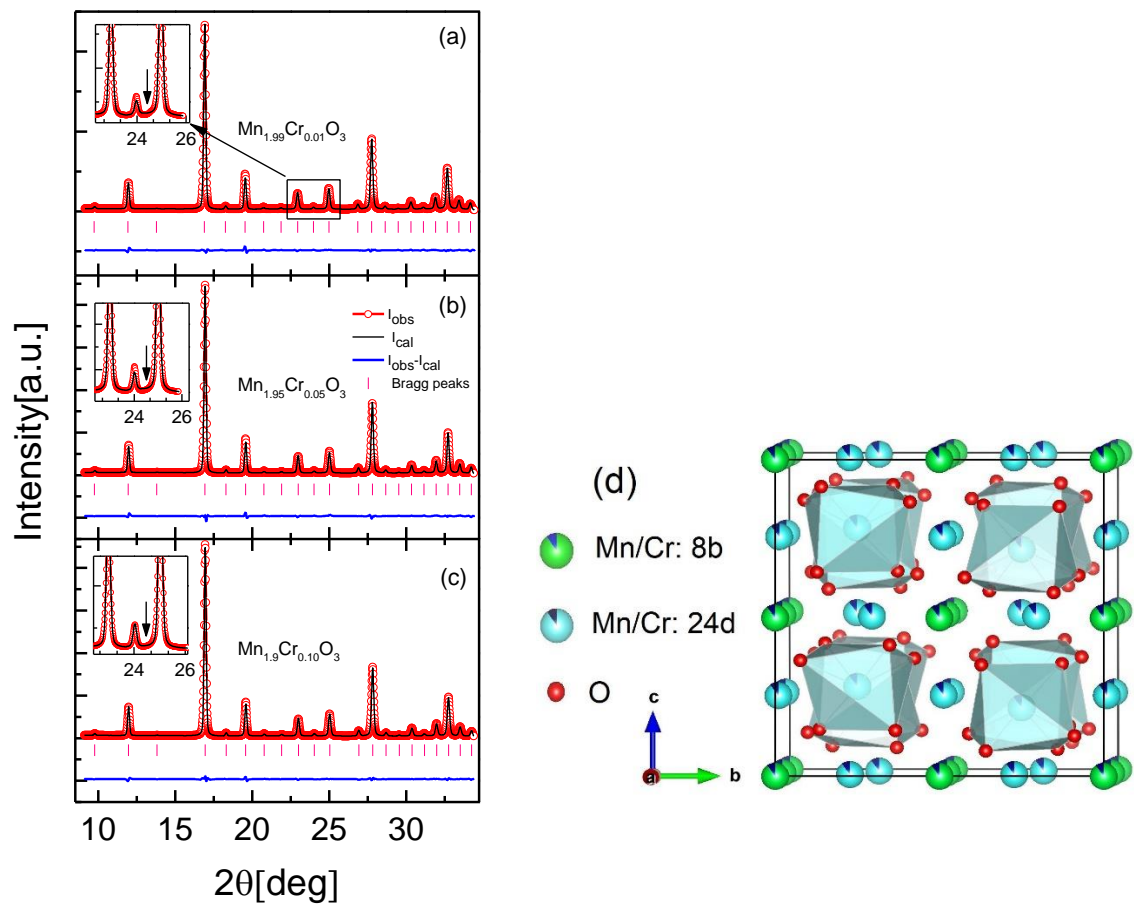

**Fig. 1**

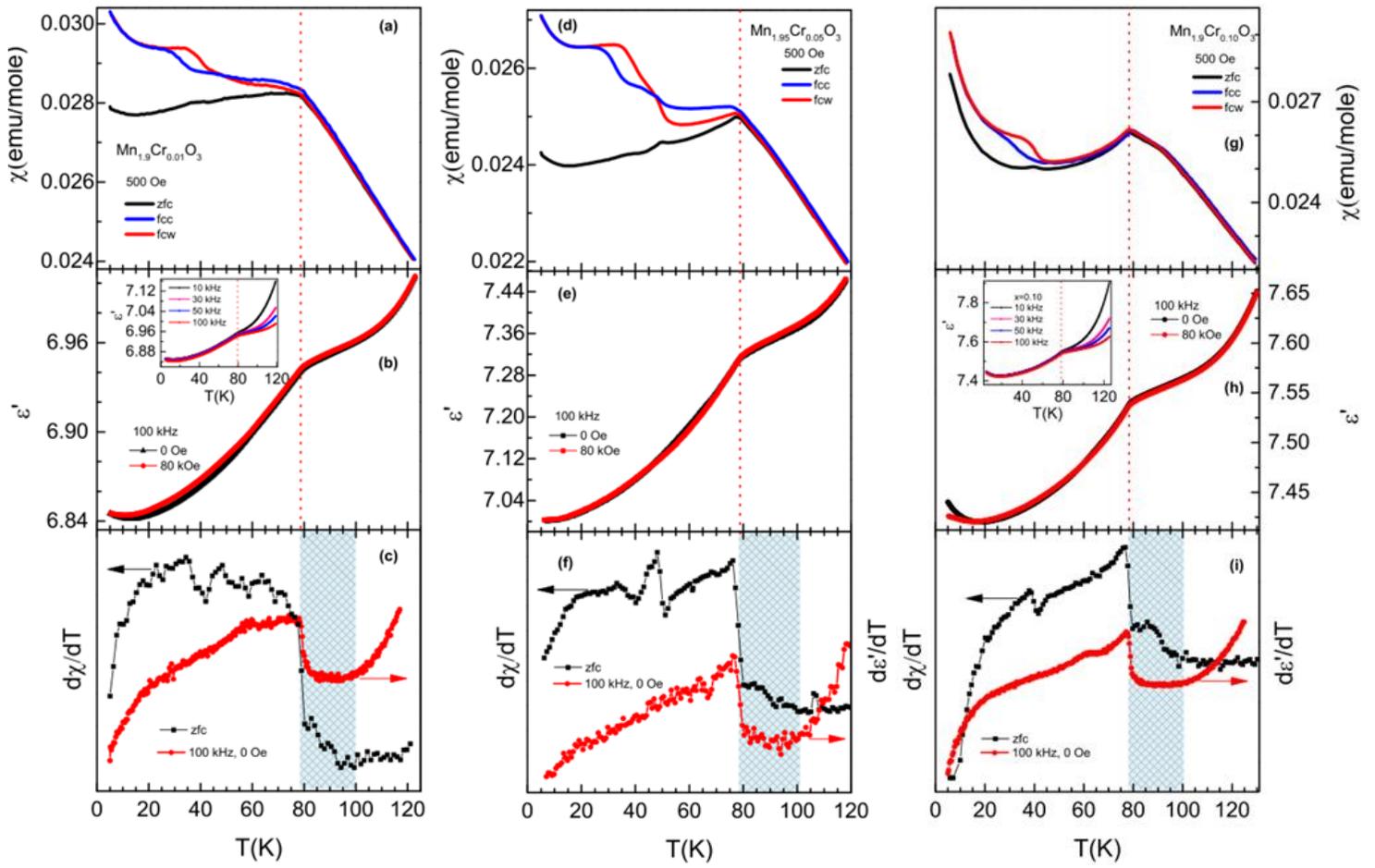

**Fig. 2**

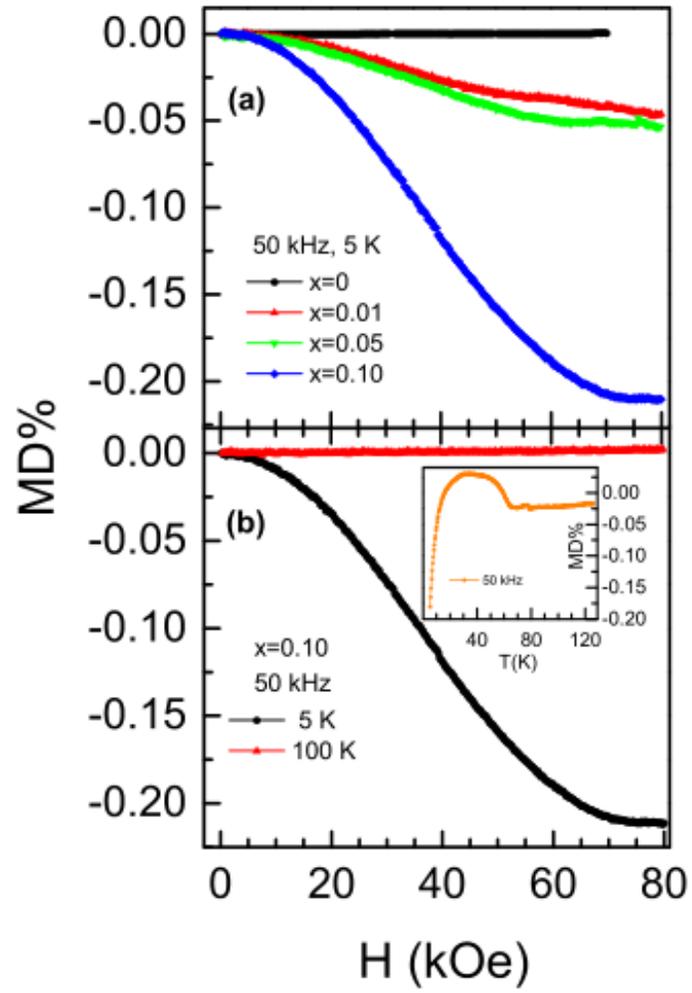

**Fig. 3**

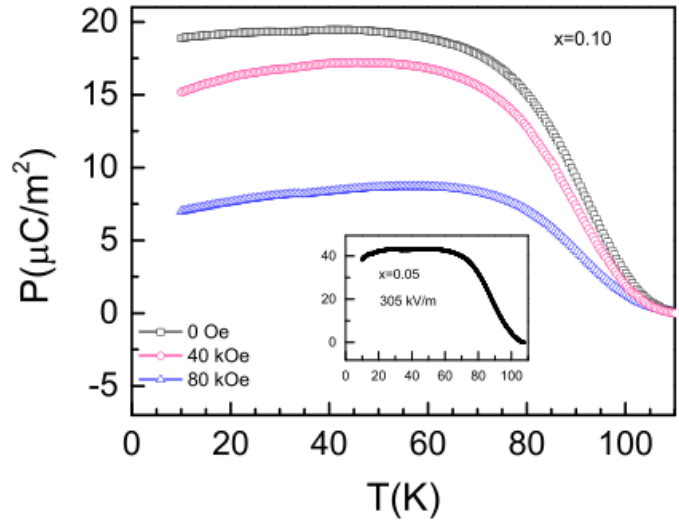

**Fig. 4**